\documentclass[aps,prd,twocolumn,showpacs,nofootinbib]{revtex4-1}  
\usepackage{graphicx}  
\usepackage{bm}        
\usepackage{amssymb}   
\usepackage{hyperref}
\usepackage{natbib}
\usepackage{amsmath}
\usepackage{color}

\begin{document}

\title{Constraints on primordial magnetic fields from CMB distortions
in the axiverse }

\author {Hiroyuki Tashiro}
\affiliation{ 
Physics Department, Arizona State University, Tempe, Arizona 85287, USA}

\author {Joseph Silk}
\affiliation{Institut d'Astrophysique, UMR 7095 CNRS, Universit\'{e} Pierre et Marie Curie, 98bis Blvd Arago, 75014 Paris, France}
\affiliation{Department of Physics and Astronomy, The Johns Hopkins University, Homewood Campus, Baltimore MD 21218, USA}
\affiliation{Beecroft Institute of Particle Astrophysics and Cosmology, Department of Physics, University of Oxford, Oxford OX1 3RH, UK}

\author {David J. E. Marsh}
\affiliation{Perimeter Institute, 31 Caroline St N, Waterloo, ON, N2L 6B9, Canada}


\date{\today}
\begin{abstract}
Measuring spectral distortions of the cosmic microwave background (CMB)
is attracting considerable attention as a probe of high energy particle physics in the
cosmological context, since PIXIE and PRISM have recently been
proposed. In this paper, CMB distortions due to resonant conversions
between CMB photons and light axion like particles (ALPs) are
investigated, motivated by the string axiverse scenario which suggests
the presence of a plenitude of light axion particles. Since these
resonant conversions depend on the strength of primordial magnetic
fields, constraints on CMB distortions can provide an upper limit on
the product of the photon-ALP coupling constant $g$ and the comoving
strength of primordial magnetic fields $B$.  Potentially
feasible constraints from PIXIE/PRISM can set a limit $g B \lesssim 10^{-16}~{\rm GeV}^{-1} {\rm nG}$ for
ALP mass $m_\phi \lesssim 10^{-14}~{\rm eV}$.  Although this result is not a
direct constraint on $g$ and $B$, it is significantly tighter than the product of the
current upper limits on $g$ and $B$.
\end{abstract}

\maketitle

\section{introduction}

An axion is a strongly motivated particle for a dark matter candidate.
The  axion was originally introduced to solve the strong CP problem in quantum
chromodynamics (QCD)~\cite{pecceiquinn1977,Weinberg:1977ma,Wilczek:1977pj}.
Axion-like particles
(ALPs) also appear naturally in string theory~\cite{Witten:1984dg,Svrcek:2006yi}. The topological complexity
of string theory compactifications can provide a plenitude of light
ALPs spanning many orders of magnitude in mass, known as the string axiverse scenario~\cite{Arvanitaki:2009fg,acharya2010a,cicoli2012c}. In principle there is no lower limit to the ALP mass in this scenario, though the lower limit of relevance for dark matter is the Hubble scale, $H_0\approx 10^{-33}$ eV.

Recent gamma-ray data from blazars suggests 
the existence of cosmological magnetic fields stronger than $10^{-16}$G
in large voids~\cite{Neronov:1900zz,Tavecchio:2010mk}.
Such magnetic fields can be  accounted for by primordial magnetic fields,
which are generated in the early universe~(for recent reviews, see Refs.~\cite{Kandus:2010nw,Widrow:2011hs,Durrer:2013pga}).
Since ALPs generally couple with electromagnetic fields,
one can expect a conversion
between cosmic microwave background (CMB) photons and ALPs in the presence of primordial magnetic fields. 
Such a conversion produces observable distortions in the CMB spectrum~\cite{Yanagida:1987nf}.
Ref.~\cite{Mirizzi:2009nq} has studied the resonant conversion between
CMB photons and ALPs. They obtained the photon-ALP mixing constraint from
the Far Infrared Absolute Spectrophotometer (FIRAS) data of the Cosmic Background Explorer
(COBE)~\cite{Mather:1993ij,Fixsen:1996nj}.
Since the COBE FIRAS constraint on the resonant conversion probability $P$
corresponds to $P< 5 \times 10^{-5}$,
they provided a constraint $g B < 10^{-13} \sim 10^{-11} {\rm GeV}^{-1} {\rm nG}$ for ALP masses between $10^{-14}$~eV and $10^{-4}$~eV, where
$g$ is the coupling constant and $B$ is the spatially averaged magnetic field strength
at the present epoch.
Their result suggests that, if primordial magnetic fields have a
strength close to the current upper limit, the CMB distortion constraint
gives a stronger constraint on $g$ than the solar
and astrophysical bounds~\cite{Andriamonje:2007ew,Raffelt:2006cw,Friedland:2012hj}.

Recently, PIXIE~\cite{Kogut:2011xw} and PRISM~\cite{Andre:2013afa} have
been proposed to provide precision measurements of the CMB frequency
spectrum.
Measuring CMB distortions form the blackbody spectrum is a good probe to access
the thermal history of the
universe~\cite{Zeldovich:1969ff,1970Ap&SS...7...20S,
Danse:1977,1991A&A...246...49B,Hu:1992dc}~(see Refs.~\cite{Chluba:2011hw,Sunyaev:2013aoa} for recent reviews).
The current goal of
these sensitivities to CMB distortions is set to be a factor $\sim 10^4$
improvement on the  COBE FIRAS.
Here, we revisit 
CMB distortions due to resonant photon-ALP conversions, and make a forecast about the feasibility of  future
constraints from PIXIE/PRISM.
We also expand the constraint to smaller ALP masses, $<10^{-14}$~eV,
than in Ref.~\cite{Mirizzi:2009nq}.
Such small-mass ALPs naturally arise in the axiverse scenario and have
diverse phenomenology in the CMB, large scale structure, and black hole
astrophysics that can constrain them independently of their couplings to
the visible
sector~\cite{hu2000,amendola2005,marsh2013,marsh2013b,arvanitaki2011}.
However, when the coupling to photons is present such ALPs can go
through resonant conversions with CMB photons due to plasma effects in
the cosmic dark age and be constrained independently of their
contribution to dark matter.
We evaluate CMB distortions due to small-mass ALPs, taking into account multiple resonant conversions.

This paper is organized as follows. We briefly review the resonant
conversion between photons and ALPs in the cosmological scenario in
Sec.~II. We also derive the analytical form of the resonant conversion
in both
strong and weak coupling limits at the resonant epoch.
In Sec.~III, we calculate the resonant conversion probability numerically
and evaluate the PIXIE/PRISM constraints on the ALP coupling and
primordial magnetic field strength. 
Sec.~IV is devoted to our conclusions.
Throughout this paper, we adopt natural units where $\hbar=1$, $c=1$ and
the Boltzmann constant $k_B=1$.
We use cosmological parameters for a flat
$\Lambda$CDM model,
$h=0.69$, $\Omega_{\rm B} h^2 =0.022$,  and $\Omega_{\rm C} h^2 =0.11$.

\section{resonant conversions}\label{sec:conversion}

ALPs couple to electromagnetic fields through a two photon
vertex.
In the existence of external magnetic
fields, electromagnetic fields in the ALP interaction terms can be decomposed into the dynamical part of
photons and the external magnetic field part. As a result, the
interaction term of an ALP and a photon with external magnetic fields is
given by
\begin{equation}
 {\cal L} = g \omega B_T A_{||} \phi,
\end{equation}
where $\omega$ is the photon frequency, $B_T$ is the component of the external magnetic field
perpendicular to the propagation of photons, and $A_{||}$ is
the component of a photon parallel to the $B_T$ component.

Due to this interaction, the propagation eigenstates of the photon-ALP
system $(\gamma, \phi)$ are different from the interaction
eigenstates with external magnetic fields.
Therefore, conversion between $\gamma$ and $\phi$ occurs
in the same way as for massive neutrinos of
different flavors.
The mixing angle of $(\gamma, \phi)$ in vacuum is given by~\cite{Raffelt:1987im}
\begin{eqnarray}
\cos 2 \theta_{\rm v} & =&\frac{m_\phi^2}{\sqrt{(2g B_T \omega)^2 + m_\phi^4}},
\nonumber\label{eq:mix_vac} \\
\sin 2 \theta_{\rm v} & =&\frac{2g B_T \omega}{\sqrt{(2g B_T \omega)^2 +m_\phi^4}},
\label{eq:theta0}
\end{eqnarray}
where $m_\phi$ is the ALP mass.
This mixing angle produces photon-ALP
oscillations with a wavenumber 
\begin{equation}
k =  \frac{\sqrt{m_\phi^4+(2g B \omega)^2}}{2\omega} .\label{eq:wavek}
\end{equation}

In the cosmological plasma, the photon dispersion relation is modified due
to plasma effects. This modification can be parametrized by an
effective photon mass $m_\gamma$.
Among the various plasma effects,
the scatterings off free electrons and neutral atoms make
negative and positive contributions to the effective photon mass, respectively.
Due to these effects, the effective mass can be given as~\cite{Mirizzi:2009iz}
\begin{equation}
m_\gamma^2 = \omega_p^2 -2 \omega^2 (n_{\rm H}-1),
\label{eq:m_gamma}
\end{equation}
where $\omega_p$ is the plasma frequency $\omega_p^2 = 4 \pi \alpha n_e
/m_e$ with the fine structure constant $\alpha$, the electron mass $m_e$
and the free electron number density $n_e$.
The refractive index of neutral hydrogen $n_{\rm H}$ is set to $(n_{\rm H}-1)=1.36 \times 10^{-4}$ in
normal conditions~\cite{Born:1980}. 
In Eq.~(\ref{eq:m_gamma}), we ignore the contributions of helium and
magnetic fields, because these effects are negligibly small~\cite{Mirizzi:2009iz}.

\begin{figure}[t]
\includegraphics[width=8cm]{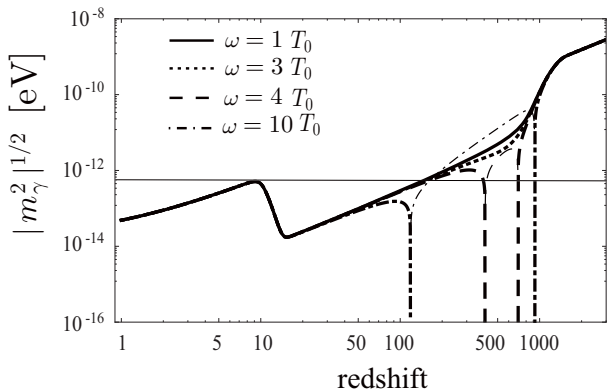}
\caption{ The evolution of the effective photon mass. Thick lines represent
 positive effective mass regions, while thin lines show negative effective
 mass regions. The solid, dotted, dashed and dashed-dotted lines are for
$\omega =T_0$, $3T_0$, $4T_0$ and $10 T_0$ with CMB temperature $T_0$.
For $\omega \lesssim 3 T_0$ the photon mass is always positive and the minimum ALP
mass that experiences resonant conversion is $m_\phi\approx
10^{-14}~$eV, with multiple resonance for $m_\phi\lesssim
10^{-12}~$eV (horizontal line). Since $m_\gamma$ passes through zero and back for high $\omega$,
these will be most relevant for light ALPs and multiple resonances.
 }
\label{fig:mg2}
\end{figure}

The effective photon mass depends on the evolution of the ionization
fraction $x(z)$ through the neutral hydrogen and free electron number
densities.
Ionized hydrogen  recombines with free electrons at $z \approx 1100$ and is
reionized around $z \approx 10$~\cite{Hinshaw:2012aka}.  
We calculate $x(z)$ with {\tt RECFAST}~\cite{Seager:1999bc},  adopting a toy model for reionization 
which is given by a $tanh$ function, $x(z) = 1+\tanh[(z-z_{re})/\Delta z] $
with $z_{re} =10$ and $\Delta z=1$. We plot the evolution of the
effective photon mass in Fig.~\ref{fig:mg2} with CMB temperature $T_0$.
The effective mass squared has positive
and negative contributions from scattering off free electrons and
neutral atoms, respectively. 
As a result, 
the effective mass becomes negative for high frequencies in the dark ages
where negative contributions dominate positive ones.

The effective photon mass modifies photon-ALP oscillations through
the Lagrangian.
This effect arises as the effective
mixing angle $\tilde \theta$~\cite{Raffelt:1996wa}, 
\begin{eqnarray}
\cos 2 \tilde \theta &=& \frac{\cos2 \theta_{\rm
 v}-\xi}{\sqrt{(\sin 2\theta_{\rm v})^2 + (\cos2 \theta_{\rm
 v}-\xi)^2}},
 \nonumber \\
 \sin 2 \tilde \theta &=& \frac{\sin 2\theta_{\rm v}}{\sqrt{(\sin
  2\theta_{\rm v})^2 + (\cos2 \theta_{\rm  v}-\xi)^2}},
\label{eq:theta}
\end{eqnarray}
where the parameter $\xi$ controls the significance of the plasma effects,
\begin{equation}
\xi  = \frac{m_\gamma^2}{\sqrt{m_\phi^4+(2 g B \omega)^2}}=
\left(\frac{m_\gamma}{m_\phi}
 \right)^2  \cos 2\theta_{\rm v}.
\label{eq:xi}
\end{equation}

As shown in Fig.~\ref{fig:mg2}, the effective mass at early universe can be much larger than the ALP mass and
$\xi \gg 1$. In this case,  the conversion between $\gamma$ and $\phi$ is suppressed.
However, as the universe evolves, the effective mass equals  an ALP mass
and $\xi$ reaches unity. At this time, the effective mixing angle becomes
$\pi/4$ and the resonant conversion occurs between photons and ALPs with this
mass.
This is in analogy to ``resonant'' neutrino oscillations
known as Mikheyev-Smirnov-Wolfenstein (MSW) effect~\cite{Wolfenstein:1977ue,Mikheev:1986gs}.

The conversion probability for the resonance is given by~\cite{Parke:1986jy}
\begin{equation}
P \approx \frac{1}{2}+\left(p-\frac{1}{2}\right)
\cos 2\theta_p \cos 2 \theta_0,
\label{eq:g-p-eq-heavy}
\end{equation}
where $p$ is the level crossing probability,
$\theta_p$ and $\theta_0$ are the effective mixing angles at the
photon production~($z \gg 1 $) and detection points~($z=0$), respectively.
At the photon production point (i.e. at reheating), since the redshift is very high, 
$\xi$ is large. Hence we approximate $\cos 2 \theta_p \sim
-1$ throughout this paper.

The level crossing probability $p$ indicates the non-adiabaticity of the conversion. 
While $p$ becomes zero in the limit of the adiabatic conversion,
$p$ reaches unity in the extremely non-adiabatic case.
In order to obtain the level crossing probability, we 
make the approximation
that $m_\gamma$ varies linearly in the resonance regime, and make a Taylor-series expansion at the resonance position,  neglecting the
second- and higher-derivative terms.
In this approximation, the level crossing probability is given by applying the Landau-Zener result~\cite{Parke:1986jy},
\begin{equation}
p \approx \exp\left[ - \frac{k_r r}{2} \frac{\sin^2 2 \theta_r }{\cos 2
	       \theta_r} \right],
\label{eq:landau}
\end{equation}
where $k_r$ and
$\theta_r$ are respectively the oscillation wavelength and vacuum
mixing angle at the resonance epoch, and
$r$ is a scale parameter to be evaluated at the location where
a resonance occurs~\cite{Mirizzi:2009iz},
\begin{equation}
r=\left|\frac{d \ln m^2_\gamma(t)}{dt}\right|^{-1}_{t=t_r},
\label{eq:r}
\end{equation}
where $t_r$ refers to the time at the resonance.

For small $m_\phi$~($m_\phi < 10^{-12}$~eV), we expected from Fig.~\ref{fig:mg2} that
multiple resonances occur. 
The conversion probability can be calculated in a manner similar to the
single resonance case. Following the classical probability result, we obtain
the probability for the double resonant case by replacing $p$
by $p_1 (1-p_2)  + p_2 (1-p_1)$, where $p_1$ and $p_2$ are the
probabilities for the first and second resonances, respectively~\cite{Parke:1986jy}.
Similarly to the double resonant case, the probability for more multiple
resonances can be calculated.

Resonant photon-ALP conversions depend on the component of external
magnetic fields perpendicular to the propagation direction of photons,
$B_T$.  Generally, primordial magnetic fields have structures which
depend on the generation mechanisms of these fields. Therefore, the
resonant conversion probability is possibly anisotropic due to these structures and depends on the ratio of the
magnetic field coherent scale to the resonant scale.
In this paper, we focus on the monopole component of CMB distortions to the black-body spectrum.
%

Before calculating resonant conversion probabilities numerically, it is worth
estimating them analytically for two cases, $m_\phi \gg m_{\gamma 0}$
and $m_\phi \ll m_{\gamma 0}$ where $m_{\gamma 0}$ is the effective
photon mass at the detection point~($z=0$).

{\sc In the case of $m_\phi \gg m_{\gamma 0}$},
the resonant conversion probability has been studied in Ref.~\cite{Mirizzi:2009nq}.
The mixing angle at the detection point is expressed
by  that in the vacuum state given by Eq.~(\ref{eq:theta0}).
Since we are interested in the weak mixing limit, $g \langle B^2 _0
\rangle^{1/2}
\omega \ll m_\phi^2$,
we can approximate $\cos 2 \theta_0 \approx 1$.
In this case, the resonant conversion happens only once.
Therefore the conversion probability is provided by
Eq.~(\ref{eq:g-p-eq-heavy}), and
the sky-averaged conversion probability can be approximated by
 \begin{equation}
  \langle P \rangle \approx 1 -p.
\label{eq:con-vac-approx}
 \end{equation}
In order to satisfy the COBE FIRAS limit, $\langle P \rangle < 5 \times
10^{-5}$,
a strong non-adiabatic resonance, $p \sim 1$, is required.
For the single resonance case, $p$ is given by Eq.~(\ref{eq:landau}).
A strong non-adiabatic condition, $p \approx1 $, leads the level
crossing probability to
\begin{equation}
 p \approx 1  - \frac{k_r r}{2} \frac{\sin^2 2 \theta_r }{\cos 2
	       \theta_r}.
\label{eq:expand}
\end{equation}
Accordingly, in the weak coupling limit, the conversion probability at a comoving frequency $\omega$
can be provided by~\cite{Mirizzi:2009nq}
\begin{equation}
\langle P \rangle
 \approx \pi r \omega   
\frac{g^2 \langle B^2_0 \rangle (1+z_r)^5}{    m_\phi^2 },
\label{eq:con-approx}
\end{equation}
where the redshift
factor comes from the dependence of both magnetic fields and CMB
physical frequency
on the resonant epoch.

For $10^{-14} ~{\rm eV} <m_\phi <5 \times 10^{-13}~{\rm eV}$,
CMB photons suffer multiple resonances.
Since photons are still detected in vacuum,
we can approximate $\cos 2 \theta_0 \approx 1$ in the weak mixing limit.
For the double resonance case,
the sky-averaged conversion probability is
expressed as
\begin{equation}
\langle P \rangle \approx 1- p_1(1-p_2) -p_2 (1-p_1).\label{eq:tuned}
\end{equation}
It is clear that the COBE FIRAS limit requires fine-tuning of $p_1$ and $p_2$.
Accordingly, for $10^{-14} ~{\rm eV} <m_\phi <5 \times 10^{-13}~{\rm eV}$,
the parameter regions of $g$ and $\langle B^2 \rangle^{1/2}$ are tightly restricted.

{\sc In the case of $m_\phi \ll m_{\gamma 0} $},
photons are no longer detected in vacuum, and the mixing angle at the
detection point is given
by Eq.~(\ref{eq:theta}). Therefore, $\cos 2 \theta_0$ is
approximated by $\cos 2 \theta_0 \sim -1$ in the weak mixing limit.
Then, the sky-averaged probability for the resonant conversion can be written as
\begin{equation}
\langle P \rangle \approx p.
\end{equation}

For low frequencies ($\omega < 4 T_0$),
the effective photon mass
is always positive and larger than the ALP mass.
Therefore, there is no resonance and
the photon-ALP conversion
is adiabatic.
This results in the sky-averaged probability for the resonant conversion
being $\langle P \rangle =0$ with $p=0$.
However, for high frequencies~($\omega \gtrsim 4 T_0$),
double resonant conversions happen. 
The sky-averaged probability for the resonant conversion is given by
\begin{equation}
\langle P \rangle \approx p_1 (1-p_2) + p_2 (1-p_1).
 \end{equation}
In the double resonant case, the COBE FIRAS limit requires 
strong non-adiabatic resonances, $p_1  \sim 1$ and $p_2  \sim 1$,
or completely adiabatic conversions, $p_1 \ll 1$ and $p_2 \ll 1$. 
Here, we focus on only the case of strong non-adiabatic resonances.
When the weak coupling limit is valid at the resonant epochs,
the resonant conversion probability is given by
\begin{equation}
 \langle P \rangle \approx \pi \omega   
\frac{g^2 \langle B^2_0 \rangle }{    m_\phi^2}
\left[r_1 (1+z_1)^5  + r_2 (1+z_2)^5
\right],\label{eq:weak-double}
\end{equation}
where the subscripts $1$ and $2$ denote values at the first and second
resonant epochs, respectively.

The magnetic field strength and CMB frequency at the resonant epochs depend
on the redshift due to the cosmological expansion. As $m_\phi$
decreases,
the weak coupling limit is no longer valid
and the strong coupling limit $g \langle B_0^2 \rangle^{1/2}  \omega \gg m_\phi^2$ is satisfied at the resonant epochs, although
the weak coupling limit
is still valid at the detection point~($z=0$).
We find that the level crossing probability in the strong coupling limit
has the same approximate form as in the weak coupling limit~(see also Appendix).
Accordingly, the sky averaged conversion probability can be written as
\begin{equation}
  \langle P \rangle \approx \pi \omega   
\frac{g^2 \langle B^2_0 \rangle }{    m_\phi^2}
\left[r_1 (1+z_1)^5  + r_2 (1+z_2)^5
\right]. \label{eq:double-reso}
\end{equation}
In both weak and strong coupling limits, 
the resonant conversion probability can be expressed in the same form
as shown in Eqs.~(\ref{eq:weak-double}) and (\ref{eq:double-reso}).

\section{CMB distortions and constraints}

A photon-ALP coupling causes resonant conversions of CMB photons to
ALPs. In the standard Big Bang scenario, the CMB spectrum is predicted as
a black-body spectrum.
When conversions happen after the decoupling of CMB thermal
equilibrium, they can produce observable CMB distortions from the
black-body spectrum.

COBE FIRAS confirmed that
the CMB spectrum is well fit to a black-body spectrum at temperature
$T_0 = 2.725 \pm 0.002~$K~\cite{Mather:1993ij}. The intensity of the black-body spectrum is 
given by
\begin{equation}
 I_0 (\omega) = \frac{\omega ^3 }{2 \pi^2} \left[ \exp(-\omega /T(z)) -1\right]^{-1}.
\end{equation}

Due to the presence of photon-ALP conversions, the observed intensity is
modified to~\footnote{This equation is valid for ALP mass  resonances
occurring after the recombination epoch. When the resonant conversion happens before
the recombination epoch, the distortion produced by the resonant
conversion is modified, because CMB photons still couple with cosmic
plasma. For a detailed analysis of this case, see Ref.~\cite{Mirizzi:2009nq}}
\begin{equation}
 I _{\rm obs}(\omega) =  I_0 (\omega) (1-P (\omega)).
\end{equation}
Therefore, the sky-average CMB distortion from the black-body spectrum can be written as
\begin{equation}
 \frac{\Delta I (\omega ) }{I_0(\omega)} =\frac{\langle I_{\rm obs} (\omega) \rangle
-  I_0(\omega)}{I_0(\omega)} = -\langle P (\omega )\rangle.
\end{equation}

\begin{figure}[t]
\includegraphics[width=8cm]{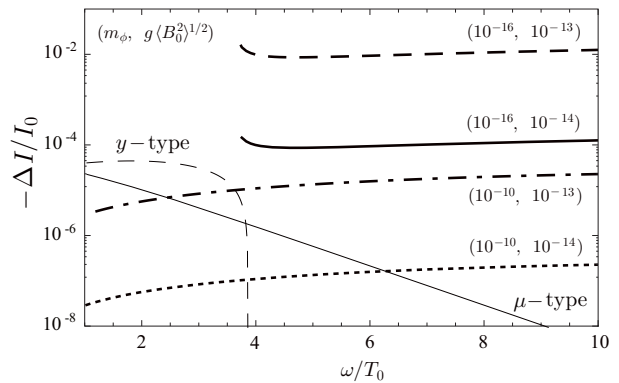}
\caption{ Frequency dependence of the resonant conversion probability.
The solid and dashed lines represent $g \langle B^2_0 \rangle ^{1/2} =
 10^{-14}~{\rm GeV}^{-1} {\rm nG}$ and $10^{-13}~{\rm GeV}^{-1}
 {\rm nG}$ for 
 $m_\phi = 10^{-16}$~eV, respectively.
The dashed-dotted and dotted lines are for 
 $g \langle B^2_0 \rangle ^{1/2} =
 10^{-14}~{\rm GeV}^{-1} {\rm nG}$ and $10^{-13}~{\rm GeV}^{-1}
 {\rm nG}$ for 
 $m_\phi = 10^{-10}$~eV, respectively.
For comparison, we plot $\mu$- and $y$-type distortions with COBE limits
in thin solid and dashed lines, respectively.
Resonance for masses $m \lesssim 10^{-14}$ eV only occurs for $\omega \gtrsim 4 T_0$, as can be seen also in Fig. 1. }
\label{fig:fr}
\end{figure}

 
We show the frequency dependence of the CMB distortions for different
ALP masses and $g \langle B^2 _0\rangle^{1/2}$ in Fig.~\ref{fig:fr}.
Note that, since the resonant conversion creates the deficit from the
black-body spectrum, the y-axis denotes $-\Delta I/ I_0$.
In this figure, we use the frequency normalized to the CMB temperature,
$T_{0}$, and normalize $m_\phi$ and $g \langle B^2_0\rangle
^{1/2}$ in units of eV and ${\rm GeV}^{-1} $nG, respectively.
For $m_\phi = 10^{-10}$~eV, the resonant conversion happens only once.
Both lines for $m_\phi = 10^{-10}$~eV in Fig.~\ref{fig:fr}
shows that the conversion probability depends on $g^2 \langle B^2 _0\rangle$ , and
is proportional to $\omega$. These are well described by Eq.~(\ref{eq:con-approx}).

On the other hand, for $m_\phi = 10^{-16}$~eV, the number of resonant
conversions depends on the CMB frequencies.
While there is no resonant conversion at low frequencies $x <4$, 
the resonance conversion occurs twice at high frequencies $x \gtrsim 4$. 
At both resonant epochs ($z>100$), the strong coupling condition is satisfied
for $g \langle B^2 _0 \rangle ^{1/2} \gtrsim 10^{-4}$. Therefore, 
the resonant conversion probability can be written as Eq.~(\ref{eq:double-reso}).
In Fig.~\ref{fig:fr},
both lines for $m_\phi = 10^{-16}$~eV show that the resonant conversion probability
also depends on $g \langle B^2 _0\rangle^{1/2}$ as expected from
Eq.~(\ref{eq:double-reso}).
The frequency dependence is not trivial, because the resonance epochs
depend on the frequency as shown in Fig.~\ref{fig:mg2}. When we assume
$z_1 > z_2$,
we find that, although the magnetic field strength and CMB frequency are
larger at higher redshift, the second term in
Eq.~(\ref{eq:double-reso}) makes a dominant contribution.
Fig.~\ref{fig:fr} shows that, as the CMB frequency
increases, $\langle P \rangle$ also becomes large on the high frequency
side $\omega \gtrsim 5 T_0$, similarly to the case
of $m_\phi> 10^{-10}$~eV where the conversion probability is proportional to
the frequency.

CMB distortions are also created in the thermalization process of energy
injections in the early universe. Ussually, these distortions are 
described by two types of the distortions; $\mu$- and $y$-type distortions~\cite{1975SvA....18..691I}.
For comparison, in Fig.~\ref{fig:fr}, we show the frequency dependence of 
$\mu$- and $y$-type distortions. Here, we plot these distortions with
COBE limits, $\mu < -1.5 \times 10^{-5}$ and $y < 9. \times 10^{-5}$~\cite{Fixsen:1996nj}.
Compared with these distortions,
the resonant conversion creates frequency-independent distortions
makes differences on high frequencies. 
Although $\mu$-type distortions are suppressed and $y$-type provides 
excess distortions ($\Delta I/I_0 >0$), the resonant conversion generates
the deficit distortions.

\begin{figure}[t]
\includegraphics[width=8cm]{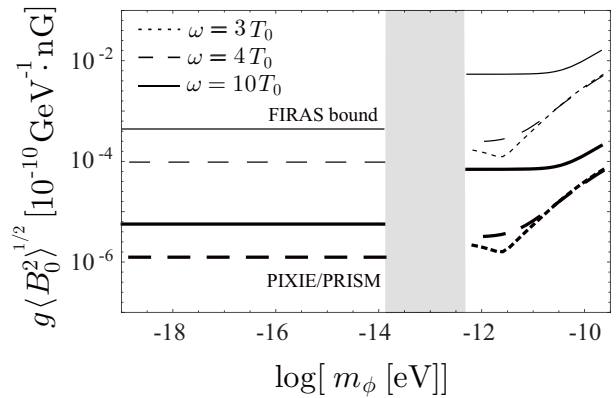}
\caption{ Constraint on $g \langle B^2_0 \rangle ^{1/2}$ as functions of
 ALP mass $m_\phi$.
The thick lines are for PIXIE/PRISM limit and the thin lines are for
 COBE FIRAS limit.
The dotted, dashed and solid lines represent frequencies, $\omega
 =3T_0$, $\omega=4 T_0$
and $\omega =10 T_0$, respectively.
The mass region where fine tuned $g \langle B^2_0 \rangle ^{1/2}$ is
 required to satisfy the COBE FIRAS limit is shown as the shaded region.
 }
\label{fig:limit}
\end{figure}

Currently, PRISM is designed to measure the deviation $\Delta I_{\rm PRISM} (\omega)
\approx 2 \times 10^{-26}~{\rm Wm^{-2} Hz^{-1} Sr^{-1}}$
for the CMB frequency range, $30~{\rm GHz} ~< \omega < 1000~{\rm GHz}$~($0.5
< x <17.5$)~\cite{Andre:2013afa}.
We adopt these values as the PIXIE/PRISM sensitivity for CMB
distortions.
We evaluate the PIXIE/PRISM constraints on the ALP coupling and magnetic
fields, $g \langle B^2 _0 \rangle ^{1/2}$ from these values.
We plot the PIXIE/PRISM constraint 
as functions of an ALP mass $m_\phi$ in Fig.~\ref{fig:limit}.

COBE/ FIRAS has provided the best upper bound on possible CMB distortions from the
black-body spectrum.
According to Ref.~\cite{Fixsen:1996nj}, we assume that COBE FIRAS
constraint is $\Delta I_{\rm FIRAS}
(\omega) <  3 \times 10^{-22} ~{\rm Wm^{-2} Hz^{-1} Sr^{-1}}$ for
frequencies $T_0 < \omega <10 T_0$ and evaluate the COBE FIRAS limit.
For comparison, we show the COBE FIRAS limit as thin lines in Fig.~\ref{fig:limit}.

For $m_\phi > 10^{-12}$~eV, the resonant conversion occurs only once.
The conversion probability can be written as Eq.~(\ref{eq:con-approx})
and is proportional to $g^2 \langle B^2 _0 \rangle$.
Therefore, as shown in Fig.~\ref{fig:limit},
PIXIE/PRISM can give a much better constraint on $g \langle B^2 _0
\rangle^{1/2}$ than COBE FIRAS and the improvement is roughly
given by the ratio $\sqrt{ \Delta I_{\rm PRISM}
/\Delta I_{\rm FIRAS} }\sim 10^2$.

In the shaded region, the multiple resonances happen while the detection
point ($z=0$) is in vacuum, $m_\phi \gg m_{\gamma 0}$. The resonant conversion probability is provided by
Eq.~(\ref{eq:tuned}). As discussed above, the COBE constraint requires 
fine tuned $g \langle B^2 _0\rangle^{1/2}$. This means that ALPs in
this mass region are effectively ruled out.

In the case of $m_\phi < 10^{-14}$~eV, there is no resonant conversion
for lower frequencies $\omega < 4 T_0$. The strong constraint on $g \langle B^2 _0
\rangle ^{1/2}$ cannot be obtained for such low frequencies.
On the other hand, CMB photons with higher frequencies $\omega \gtrsim 4
T_0$ suffer double resonant conversions.
As a result, the constraints on $g \langle B^2 _0
\rangle^{1/2} $ are divided to three ALP mass regions which correspond to
Eqs.~(\ref{eq:weak-double}) and (\ref{eq:double-reso}).

When $g \langle B^2 _0 \rangle ^{1/2}$ attains the value of the PRISM/PIXIE
constraint in Fig.~\ref{fig:limit}, the weak coupling limit is valid at resonance epochs for $m
\gtrsim 10^{-17}~$eV. 
The resonant conversion probability is given by Eq.~(\ref{eq:weak-double}).
Although Eq.~(\ref{eq:weak-double}) has an explicit dependence on $m_\phi$,
 this dependence is canceled by $r_1$ and $r_2$
which are proportional to $m_\phi ^{-2}$ from Eq.~(\ref{eq:r})~(note that $d m_\gamma /dt$ is
independent
of $m_\phi$ in this mass region). As a result, the resonant
conversion probability does not depend on $m_\phi$.

For $m_\phi \lesssim 10^{-17}$~eV, $g \langle B^2 _0 \rangle ^{1/2}$ at the
PRISM/PIXIE constraint reaches the strong coupling limit at resonance epochs. The
conversion probabilities in the strong coupling limit
are approximated to the identical form to the weak coupling limit as shown in Eqs.~(\ref{eq:weak-double}) and (\ref{eq:double-reso}). 
The conversion probability does not depend on ALP mass, $m_\phi$.
Therefore, our constraint can be extended down to $m_{\phi} \sim
10^{-24}$~eV axions which can play an interesting role
in the context of large-scale structure formation \cite{marsh2013b}.

When the ALP mass is $m_\phi < 10^{-14}$, the resonant
conversion probability in the both limits of weak and strong couplings
is proportional to $g^2 \langle B^2 _0 \rangle$.
Therefore, PIXIE/PRISM can improve the constraint on $g \langle B^2 _0
\rangle^{1/2}$ from COBE FIRAS by the ratio $\sqrt{ \Delta I_{\rm PRISM}
/\Delta I_{\rm FIRAS} }\sim 10^2$ as same as for larger mass of ALPs,
$m_\phi \gtrsim 10^{-14}$. 

To close this section we comment on our assumption about the coherent scale of magnetic fields.
In our calculations, for simplicity, we have assumed that the coherent
scale of magnetic fields
is larger than a resonant scale. Here we discuss the validity of this assumption.

Mirizzi et al. have studied the condition for this assumption with $m_\phi
> m_{\gamma 0}$~\cite{Mirizzi:2009nq}.
The typical comoving width of a resonance $L_{rc}$ is given by $L_{rc} =(1+z_r) r \sin 2
\theta_r$.
Comparing with the comoving coherent scale of magnetic fields, $L_{Mc}$,
they showed that the condition of $L_{Mc} > L_{rc}$ is valid when the coupling
$g \langle B^2_0 \rangle^{1/2} $ satisfy
\begin{eqnarray}
 g \langle B^2_0 \rangle^{1/2} &<& 0.06 
  \left( \frac{m_\phi }{10^{-12 } ~{\rm eV}}\right )^{1/3}
\nonumber \\
 && \quad \times
\left(
 \frac{T_0}{\omega}  \right ) \left ( \frac{L_{Mc} }{1 {\rm Mpc} }\right)
  {\rm GeV^{-1}   n G}.
\end{eqnarray}
Our constraint meets this condition with the comoving coherent scale
$L_{Mc} \gtrsim 1$~Mpc.

We can also check the validity of the assumption for $m_\phi < m_{\gamma
0}$. Although resonances occur twice at $z_1 $ and $z_2$,
the lower-redshift resonance has a typical comoving width $L_{2c} =
(1+z_2) r_2 \sin 2 \theta_2 $ larger than the higher-redshift resonance ($z_1 > z_2$).
For CMB frequency $4< \omega/T_0 <10$, we get the fitting formula of
$r_2$ as 
\begin{equation}
r_2 \approx 5.7 \times 10^{-7} \left( \frac{m_\phi }{10^{-14 } ~{\rm eV}}\right)^2
  \left (-2 + \frac{ \omega }{T_0} \right )^4~\rm Mpc.
\end{equation}
Fig.~\ref{fig:mg2} shows $z_2 < 500$. Therefore, the condition $L_{Mc} >
L_{2c}$ for $4 <\omega /T_0 <10$ is achieved when the comoving coherent scale of magnetic fields satisfies 
\begin{equation}
 L_{Mc} > 1.16 \left( \frac{m_\phi }{10^{-14 } ~{\rm eV}}\right)^2 {\rm Mpc}
\end{equation}
where we used $L_{Mc} > (1+z_2) r_2$ with $z_2 <500$ and $\omega /T<
10$. As the ALP mass decreases, our constraint is valid for small-scale magnetic fields.
As the ALP mass decreases, our constraint is valid for small scale
magnetic fields, such as those expected to be produced in the QCD or
electroweak phase transitions ($L_Mc \sim 100 {\rm pc} - 10 {\rm kpc}$)~\cite{Banerjee:2003xk,Kahniashvili:2012uj}.

\section{conclusion}

In this paper, we have studied resonant conversions from CMB
photons to small-mass ALPs with primordial magnetic fields.
Resonant conversions of CMB photons
to such small-mass ALPs occur due to plasma
effects in the dark age. 
These resonant conversions can produce observable CMB distortions.
We have evaluated the resonant conversion probability between CMB photons and
ALPs, following the method of Ref.~\cite{Mirizzi:2009nq}.
In particular, we have focused on low-mass ALPs, $m_\phi < 10^{-10}$~eV.
Depending on CMB frequency, CMB photons suffer multiple resonant conversions 
to such low-mass ALPs.

Using COBE FIRAS bound on the possible CMB distortions from the
black-body spectrum, we have obtained the upper limit on $g \langle
B_0^2 \rangle^{1/2}$ where $g$ and $\langle B_0 ^2 \rangle^{1/2}$ are
the photon-ALP coupling constant and the averaged primordial magnetic
field strength at the present epoch.
Our limit is $g\langle B_0 ^2 \rangle ^{1/2} \lesssim 10^{-14}~{\rm
GeV}^{-1} {\rm nG}$ for $m_\phi \lesssim 10^{-14}~{\rm eV}$ at the CMB frequency $\omega = 4 T_0$.
We have found that the COBE constraint requires 
fine tuned $g \langle B^2 _0\rangle^{1/2}$ for $10^{-14} ~{\rm eV}
\lesssim m_\phi \lesssim 4 \times 10^{-13}~{\rm eV}$. This means that ALPs in
this mass region with non-zero photon coupling are effectively ruled out.

We have also evaluated the constraint expected from PIXIE/PRISM, one of whose goals
is to measure the CMB frequency spectrum precisely.  Compared with COBE
FIRAS, PIXIE/PRISM can improve the constraint on $g \langle B_0^2
\rangle^{1/2}$ by the ratio of PIXIE/PRISM's CMB distortion
sensitivity to COBE FIRAS's one.  Accordingly, the PIXIE/PRISM
constraint is $g\langle B_0 ^2 \rangle ^{1/2} \lesssim 10^{-16}~{\rm
GeV}^{-1} {\rm nG}$ for $m_\phi \lesssim 10^{-14}~{\rm eV}$ at the CMB
frequency $\omega = 4 T_0$.

The resonant conversion creates the deficit distortion. This fact means that 
the effective temperate of residual free electron gas, which is
thermally equivalent with CMB temperature before the resonant
conversion, becomes higher than the effective CMB temperature after the conversion.
As a result, in order to recover the thermal equilibrium state,
electrons with high momenta cool by Compton scatterings and ones with low
momenta absorb CMB photons. 
This signature can appear on the CMB spectrum at low frequencies~(less
than 1~GHz). 
We will address this issue in the near future.

The current direct constraint on the photon-ALP coupling for small mass
is $g \lesssim 10^{-10}~{\rm GeV}^{-1}$ for $m_{\phi} \lesssim0.01 ~{\rm eV}$~\cite{Andriamonje:2007ew}. 
Our constraint is to the product of the photon-ALP
coupling constant $g$ and the magnetic field strength $\langle B^2_0
\rangle ^{1/2}$.
The constraint on $g \langle B^2_0 \rangle^{1/2}$ has been also derived
from studying the effect of photon-ALP conversions on CMB
polarization~\cite{Tiwari:2012ig}. Our result shows that COBE FIRAS constraint is comparable
with this bound and PIXIE/PRISM can provide the better bound. 
Recently, the limit similar to the PIXIE/PRISM constraint, $g \langle B^2_0 \rangle^{1/2} \lesssim 10^{-16} ~{\rm
GeV}^{-1} {\rm nG}$, has been suggested as due to the conversion {\it
from} the axionic dark radiation {\it to} CMB photons~\cite{Higaki:2013qka}. However,
our work relies on the reverse process, which does not require the prior
production of axionic dark radiation. Therefore PIXIE/PRISM can
give the more robust constraint.

As for primordial magnetic fields, the upper bounds on primordial magnetic fields are provided
by observations of CMB anisotropies~\cite{Paoletti:2012bb,Ade:2013zuv}
and large scale structures~\cite{Pandey:2012ss, Kahniashvili:2012dy}.
The constraint on comoving magnetic field strength with the $1$~Mpc
coherent scale is 
below several nano Gauss.
Therefore, if primordial magnetic fields are detected with the upper limit strength,
the PIXIE/PRISM limit can give a stronger constraint on the photon-ALP
coupling $g$ stronger by roughly 5 orders of magnitude compared to the present limit for
small-ALP mass, $m_\phi <10^{-14}$~eV~\footnote{If the ALP is extremely light, $m_\phi \lesssim 10^{-28}$~eV,
then the photon coupling leads to cosmological birefringence
(e.g. \cite{carroll1990}) and causes rotation of CMB
polarisation. Limits from WMAP9 \cite{hinshaw2012} translate to
constraints on $g$ of a similar order of magnitude to those obtainable
by PIXIE/PRISM for intermediate scale axion decay constants $f_a\approx
10^{12}$~GeV. Planck polarisation results should improve this further.}.


On the other hand, if in future the existence of small-mass ALPs were confirmed, for example by direct detection \cite{jaeckel2010,ringwald2012}, and
the photon-ALP coupling constant was measured then the constraint of PIXIE/PRISM
constrains primordial magnetic fields.
The measurement of CMB distortions by PIXIE/PRISM can then give the stronger
constraint on the primordial magnetic fields than other current upper limits. 


%

\acknowledgments
We thank Jens Chlubas for useful comments.
HT is supported by the DOE.
The research of JS has been supported at IAP by  the ERC project  267117 (DARK) hosted by Universit\'e Pierre et Marie Curie - Paris 6   and at JHU by NSF grant
OIA-1124403. The research of DJEM was supported by Perimeter Institute for Theoretical Physics. Research at Perimeter Institute is supported by the Government of Canada through Industry Canada and by the Province of Ontario through the Ministry of Research and Innovation.

\appendix*
\section{Resonant conversion probability in the weak and strong coupling limits}\label{append}

Depending on the ALP mass, we have obtained the analytical approximations of the resonant conversion
probability in the weak coupling limit, $m_\phi^2 > g B \omega$, and the
strong coupling limit,  $m_\phi^2 < g B \omega$ in Sec.~\ref{sec:conversion}.
In particular for $m_\phi <10^{-14}~$eV, these forms are identical as shown in
Eq.~(\ref{eq:weak-double}) and (\ref{eq:double-reso}).
In this appendix, we derive these approximate forms
in the limits of weak coupling~(WC) and of strong coupling~(SC).

In both limits, the mixing angle in vacuum, Eq.~(\ref{eq:mix_vac}),
is approximated by
\begin{eqnarray}
 \cos 2 \theta_{\rm v} &\approx&
  \begin{cases}
    1 - \displaystyle\frac{1 }{ 2} \left( \frac{2 gB \omega }{
   m_\phi^2}\right)^2 \quad &{\rm WC~ limit}, \\[8pt]
   \displaystyle\frac{m_\phi^2 }{ 2 gB \omega} \quad &{\rm SC~ limit},
  \end{cases}
\\[6pt]
 \sin 2 \theta_{\rm v} &\approx&
\begin{cases}
\displaystyle\frac{2 gB \omega }{ m_\phi^2} \quad &{\rm WC~ limit},
\\[8pt]
1- \left(\displaystyle\frac
 {m_\phi^2 }{ 2 gB \omega}\right)^2 \quad &{\rm SC~ limit}.
\end{cases}
\label{eq:ap-thv}
\end{eqnarray}

The wavenumber of photon-ALP oscillations induced by this mixing angle
is expressed in these limits as
\begin{equation}
 k \approx
\begin{cases}
\displaystyle\frac{m_\phi ^2 }{ 2 \omega} \quad &{\rm WC~ limit},
\\[8pt]
 gB \quad &{\rm SC~ limit}.
\end{cases}
\label{eq:ap-k}
\end{equation}

Due to cosmic plasma effects, the effective photon mass plays a role. The
effective photon mass modifies the mixing angle.
The level crossing probability is given by Eq.~(\ref{eq:landau}).
In this paper, we focus on the limit of non-adiabatic resonance, $p \sim
1$. Therefore, the level crossing probability can be expanded as
Eq.~(\ref{eq:expand}). Accordingly, using Eq.~(\ref{eq:ap-thv}) and
(\ref{eq:ap-k}), we can approximate $p$ in the both limits of the weak and strong
coupling to 
\begin{equation}
1-p \approx
\begin{cases}
\pi r \omega   
{g^2 B^2/ m_\phi^2 } \quad &{\rm WC~ limit},
\\[8pt]
 \pi r \omega   {g^2 B^2/ m_\phi^2 } \quad &{\rm SC~ limit}.
\end{cases}\label{eq:ap-p}
\end{equation}
In both limits, the level crossing probabilities are identical.

For $m_\phi < 10^{-14}$~eV, the double resonant conversion for CMB
photons with $\omega >4 T_0$ happens.
The double resonant conversion probability is given by
\begin{equation}
P \approx \frac{1}{2}+
\left(p_1(1-p_2 )+ p_2(1-p_1)-\frac{1}{2}\right)
\cos 2 \theta_p \cos 2 \tilde \theta_0 .\label{eq:ap-double}
\end{equation}

Let us evaluate the double resonant conversion probability with $\cos 2 \theta_p =-1$.
When the ALP mass is $m_\phi < 10^{-14}$~eV, $m_{\gamma 0} \gg m_\phi$ is
satisfied as shown in Fig.~\ref{fig:mg2}.
Hence the effective mixing angle $\cos 2 \tilde \theta_0$ at the
detection point ($z=0$) can be approximated by
\begin{eqnarray}
\cos 2 \tilde \theta_0
&\approx& -1 +\frac{1}{2} \frac{\sin^2 2 \theta_{\rm v}}{ \xi^2 }
\nonumber\label{eq:ap-tildetheta} \\
&\approx&
\begin{cases}
 -1 +\displaystyle\frac{1 }{ 2} \left(\displaystyle\frac{m_\phi}{m_{\gamma0}}
 \right)^4 
 \left( \frac{2 gB_0 \omega }{
   m_\phi^2}\right)^2 
\quad &{\rm WC~ limit},
\\[10pt]
 -1 +\displaystyle\frac{1 }{ 2} 
  \left(\displaystyle\frac
 { 2 gB_0 \omega }{ m_{\gamma 0}^2}
 \right)^2 \quad &{\rm SC~limit}.
\end{cases}
\end{eqnarray}
where $B_0$ is the comoving magnetic field strength.
The current constraints on $g$ and $B_0$ are $g\lesssim 10^{-10}~{\rm
GeV}^{-1}$ and $B \lesssim 1 $~nG.
These constraints suggest that ${ m_{\gamma 0}^2} \gg gB \omega$ for $\omega
< 10 T_0$. Therefore, we can assume $\cos 2 \tilde \theta_0 \approx -1$ 
in both limits.

Plugging Eqs.~(\ref{eq:ap-p}) and (\ref{eq:ap-tildetheta}) into Eq.~(\ref{eq:ap-double}),
we obtain in the limits of weak and of strong couplings,
\begin{equation}
 P \approx \pi \omega   
\frac{g^2 B_0^2}{    m_\phi^2}
\left[r_1 (1+z_1)^5  + r_2 (1+z_2)^5
\right],
\label{eq:ap-dob}
\end{equation}
where we use $B_i =B_0 (1+z_i)^2$ and $\omega_i =\omega (1+z_i)$ with
the subscript $i$ denoting 1 and 2.
Eq.~(\ref{eq:ap-dob}) corresponds to Eqs.(\ref{eq:weak-double}) and (\ref{eq:double-reso}).

%


\end{document}